\documentclass[10pt,conference]{IEEEtran}
\usepackage{amsmath,amsfonts}
\usepackage{algorithmic}
\usepackage{graphicx}
\usepackage{textcomp}
\usepackage[usenames,dvipsnames,table]{xcolor}
\usepackage[hidelinks]{hyperref}
\usepackage[numbers,sort&compress]{natbib}
\usepackage{colortbl}

\definecolor{headercolor}{RGB}{248, 196, 145}
\definecolor{rowlight}{RGB}{255, 243, 230}
\definecolor{rowwhite}{RGB}{255, 255, 255}
\definecolor{bestred}{RGB}{210, 45, 45}
\definecolor{secondblue}{RGB}{40, 100, 195}
\definecolor{rulecolor}{RGB}{240, 200, 160}
\usepackage[capitalize]{cleveref} 
\Crefname{section}{Sec.}{Sects.}
\usepackage{xurl}
\usepackage{enumitem}
\setitemize{noitemsep,topsep=0pt,parsep=0pt,partopsep=0pt}
\usepackage{todonotes}
\usepackage{booktabs}
\usepackage{multirow}
\usepackage{tabularx}
\usepackage{subcaption}
\usepackage{microtype}
\usepackage{tikz}
\usetikzlibrary{shadows}
\usepackage[most]{tcolorbox}
\usepackage{pifont}

\newtcolorbox{boxK}{
  colback=tagblue!5, colframe=tagblue!40, boxrule=0.5pt, arc=3pt,
  left=6pt, right=6pt, top=4pt, bottom=4pt
}

\def\BibTeX{{\rm B\kern-.05em{\sc i\kern-.025em b}\kern-.08em
    T\kern-.1667em\lower.7ex\hbox{E}\kern-.125emX}}

\newcommand\parhead[1]{\vspace{+.02cm}\noindent\textbf{{#1.}}}
\definecolor{blue(ncs)}{rgb}{0.0, 0.53, 0.74}
\definecolor{realgreen}{RGB}{34, 120, 79}
\definecolor{simbrown}{RGB}{160, 82, 45}
\definecolor{tagblue}{RGB}{52, 101, 175}
\definecolor{taggray}{RGB}{130, 130, 130}
\definecolor{tagorange}{RGB}{200, 120, 20}
\definecolor{tagred}{RGB}{190, 50, 50}
\definecolor{missred}{RGB}{220, 60, 60}
\definecolor{tableheader}{RGB}{85, 115, 155}      
\definecolor{rowhighlight}{RGB}{232, 245, 233}   
\definecolor{rowalt}{RGB}{235, 242, 250}          

\tikzset{
  tag/.style={rounded corners=2pt, inner sep=1.5pt, outer sep=0pt,
              font=\sffamily\bfseries\tiny, minimum width=0.8cm, text=white, align=center},
  tag edit/.style={tag, fill=tagblue},
  tag sel/.style={tag, fill=taggray},
  tag cpl/.style={tag, fill=tagorange},
  tag term/.style={tag, fill=realgreen},
  tag err/.style={tag, fill=tagred},
  tag chat/.style={tag, fill=tagorange!80!red},
  op desc/.style={font=\tiny, anchor=west, text width=3.5cm},
  timeline/.style={line width=1.2pt},
  node dot/.style={circle, inner sep=1.2pt, fill=#1},
}

\definecolor{listCircle}{RGB}{100, 181, 226}    

\newcommand{\best}[1]{\textcolor{bestred}{\textbf{#1}}}
\newcommand{\second}[1]{\textcolor{secondblue}{\underline{#1}}}

\newcounter{resq}

\crefname{resq}{RQ}{RQs}
\crefformat{resq}{RQ#2#1#3}
\Crefname{resq}{RQ}{RQs}

\newcommand{\summary}[2]{%
\begin{boxK}
\small \ding{46} \textbf{#1:}
\textit{#2}
\end{boxK}
}

\newcommand{\challengename}{Challenge} 
\newcounter{challenge}
\crefformat{challenge}{#2\challengename\ #1#3}
\Crefformat{challenge}{#2\challengename\ #1#3}

\begin{document}

\title{An Empirical Study of Proactive Coding Assistants in Real-World Software Development}

\author{\IEEEauthorblockN{Lehui Li\textsuperscript{*,1,2}, Ruixuan Jia\textsuperscript{*,1}, Guo-Ye Yang\textsuperscript{2}, Jia Li\textsuperscript{$\dagger$,1}}
\IEEEauthorblockA{
		\textsuperscript{1}\textit{College of AI, Tsinghua University}, China \\
		\textsuperscript{2}\textit{Fitten Tech Co., Ltd.}, China \\
		\textsuperscript{*}Equal contribution \hspace{0.5cm}
		\textsuperscript{$\dagger$}Corresponding author
}}
\maketitle

\begin{abstract}\looseness=-1
Large language model (LLM)-based coding assistants have become increasingly capable, yet most remain reactive and provide assistance only after explicit developer instructions. Proactive coding assistants aim to predict developers' implicit intent from integrated development environment (IDE) interaction traces and repository context, thereby reducing the cognitive overhead for writing instructions and improving development efficiency. However, due to the lack of large-scale real-world developer behavior data, existing studies rely heavily on LLM-generated simulated data, whose fidelity to real-world data remains unclear. In this paper, we study this simulation-to-reality gap through large-scale real-world data collection. We collect IDE interaction traces from 1{,}246 experienced industry developers over three consecutive days, and construct paired LLM-generated simulated traces for controlled comparison. Our analysis shows that LLM-generated simulated traces differ substantially from real-world traces in behavioral diversity, temporal structure, and exploratory behavior. Based on the collected real-world traces, we build \textbf{ProCodeBench}, a benchmark for proactive intent prediction in real-world development scenarios. Experiments on representative LLM, retrieval-augmented, and agent-based baselines show that existing methods remain far from reliable under real-world IDE traces, suggesting that simulation-based evaluation may overestimate real-world intent-prediction performance. Finally, our training study shows that LLM-generated simulated data alone cannot substitute for real-world data, but can improve performance when used before real-world fine-tuning. These findings highlight the importance of real-world developer behavior data for evaluating and training proactive coding assistants, while also revealing the complementary role of LLM-generated simulated data.
\end{abstract}


\section{Introduction}
\label{sec:introduction}

\looseness=-1
Recent advances in Large Language Models (LLMs) have substantially improved their performance on software engineering tasks such as code generation~\cite{chen2021evaluating,li2022competition} and test generation~\cite{schafer2023empirical}, leading to the increasing integration of LLM-based coding assistants into modern software development workflows~\cite{sergeyuk2026evolving,puryear2022github,barke2023grounded,liang2024large,khojah2024beyond}. Their capabilities have expanded rapidly, from early code completion and refactoring~\cite{bruch2009learning,raychev2014code} to recent coding agents~\cite{yang2024sweagent} that support multi-turn interaction and autonomous task execution. Despite this progress, most coding assistants still follow a reactive interaction paradigm, providing assistance only after developers issue explicit instructions~\cite{chen2025need,zhao2025codinggenie}. This design limits their usefulness in real-world development scenarios. On the one hand, developers must continuously formulate detailed instructions during the coding process, which introduces considerable cognitive overhead~\cite{tang2026programming,kuo2026developer,mozannar2024reading}. More importantly, because software engineering tasks are often complex, developers may struggle to clearly articulate their development intent~\cite{tang2026programming}.

\begin{figure}[!t]
  \centering
  \includegraphics[width=\linewidth]{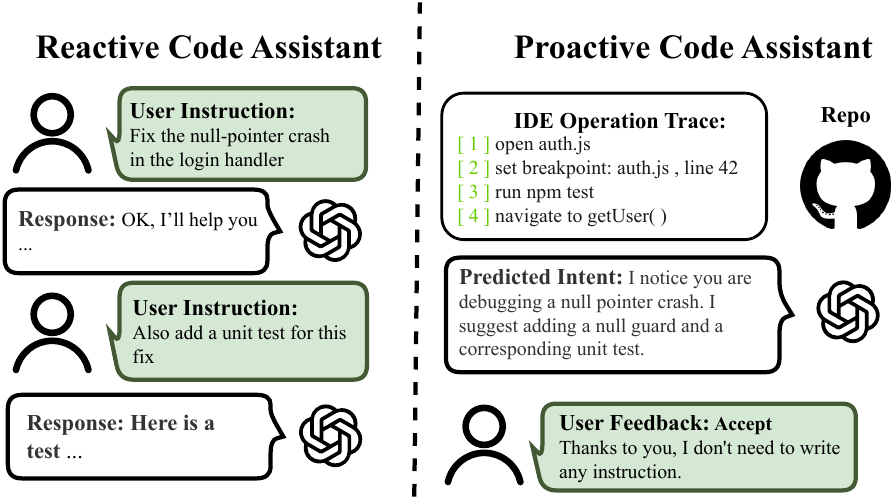}
  \caption{Comparison of reactive and proactive coding assistants. Reactive coding assistants require explicit instructions for each interaction, while proactive coding assistants predict developers' latent intent from IDE interaction traces, eliminating the need for explicit requests.}
  \label{fig:motivation}
\end{figure}

\looseness=-1
To address these limitations, recent studies have proposed proactive coding assistants~\cite{lu2024proactive,chen2025need,zhao2025codinggenie}. As illustrated in \Cref{fig:motivation}, unlike reactive coding assistants that wait for explicit instructions, proactive coding assistants infer developers' latent intent from IDE interaction traces and repository context, and then provide corresponding assistance suggestions. Previous user studies have shown that by identifying the intent embedded in developers' IDE interaction traces, proactive coding assistants can improve developers' task performance by 12\%--18\% on average, while also leading to notable gains in user experience for most participants~\cite{tang2026programming,kuo2026developer}.

\looseness=-1
However, proactive coding assistants still lack large-scale real-world data on developer behavior~\cite{lu2024proactive,chen2025need,zhao2025codinggenie}. Unlike traditional software engineering tasks, proactive code assistance requires continuous logging of developers' IDE interaction traces in real-world development scenarios~\cite{lu2024proactive}, which makes data collection costly and subject to strict privacy constraints. As a result, existing studies mostly rely on LLM-generated simulated data~\cite{lu2024proactive,kim2025propersim,chen2025need,zhao2025codinggenie}. They typically use LLM-based user agents to synthesize IDE interaction traces and generate the corresponding intent labels, which are then used to train and evaluate the intent-prediction ability of proactive coding assistants.

\looseness=-1
Although LLM-generated simulated data has enabled progress in proactive coding assistants, it remains unclear whether such data reflects how developers behave in real-world development scenarios. This motivates our first research question:
\summary{RQ1}{Can LLM-generated simulated data faithfully capture real-world developer behavior and the underlying intent?}
\noindent To answer this question, we collected over 4 million real-world IDE interaction traces through a Visual Studio Code (VS Code) extension. The data covers 1,246 volunteers over three consecutive days and spans representative development scenarios, including frontend, backend, database, and algorithm engineering. For each real-world trace, we further synthesize a paired LLM-generated simulated trace, which enables a controlled comparison between real-world and LLM-generated simulated data. Our gap analysis reveals a clear simulation-to-reality gap. Real-world traces show greater behavioral diversity, finer-grained operations, and more complex temporal patterns. They also contain richer but noisier process-level information. In contrast, LLM-generated simulated traces are easier to generate, but they often follow simplified behavioral patterns and fail to approximate real development processes.

\looseness=-1
Because simulated and real-world data differ substantially, evaluations based on simulated benchmarks may overestimate models' ability to proactively predict user intent. We therefore further examine:

\summary{RQ2}{How do existing proactive coding assistants actually perform when evaluated on real-world data?}

\noindent To answer RQ2, we construct ProCodeBench from the collected real-world IDE interaction traces. Because developer intent is implicit in continuous IDE interaction traces, we convert raw traces into standardized intent-prediction instances through an annotation pipeline. We then conduct a broad evaluation of 13 competitive baselines, covering seven current LLMs (e.g., GPT-5.4, Claude Sonnet 4.6, and Gemini 3.1 Pro), four Retrieval-Augmented LLMs (e.g., RepoCoder and RepoGraph)~\cite{zhang2023repocoder,ouyang2025repograph}, and two LLM-based Agents (SWE-Agent and A-RAG)~\cite{yang2024sweagent,du2026rag}. The results reveal three findings. \ding{182} Current baselines still struggle to predict developer intent from real-world IDE interaction traces, with performance substantially below that reported on simulation-based benchmarks~\cite{lu2024proactive,kim2025propersim}. \ding{183} Repository-level code context consistently improves intent-prediction performance across backbone models, indicating that repository information helps clarify the purpose behind observed IDE operations. \ding{184} LLM-based Agents achieve the strongest results through multi-turn tool use, but how to use repository-level code context effectively and efficiently remains an open challenge.

\looseness=-1
The poor real-world performance observed in RQ2 raises a further question: 

\summary{RQ3}{Can training on simulated or real-world data improve proactive intent prediction?}

\noindent To answer RQ3, we compare models trained with real-world data, LLM-generated simulated data, and a mixed-data training regime under a unified setting. The results show that training on LLM-generated simulated data alone does not transfer well to real-world development scenarios. However, the mixed-data training regime improves real-world performance when LLM-generated simulated data is used as an initialization before fine-tuning on real-world data. This suggests that real-world and LLM-generated simulated data are not interchangeable substitutes, but complementary data sources for improving proactive intent prediction. Our main contributions are as follows:
\begin{itemize}[leftmargin=1.5em, labelsep=0.4em]
    \item[\ding{182}] We present the first large-scale real-world dataset for proactive code assistance. The dataset contains three consecutive days of IDE interaction traces from 1,246 volunteers across representative development scenarios. By pairing each real-world trace with an LLM-generated simulated counterpart, we reveal a clear simulation-to-reality gap in developer behavior, including differences in behavioral diversity, operation granularity, temporal patterns, and process-level noise.

    \item[\ding{183}] We build \textbf{ProCodeBench}, the first benchmark that evaluates proactive code assistance in real-world development scenarios. We convert raw IDE interaction traces into standardized intent-prediction instances through an annotation pipeline, and provide a unified evaluation protocol. Experiments on mainstream LLMs, Retrieval-Augmented LLMs, and LLM-based Agents show that existing models still struggle with real-world proactive intent prediction.

    \item[\ding{184}] We analyze how real-world data, LLM-generated simulated data, and a mixed-data training regime contribute to model training. Our results show that LLM-generated simulated data alone does not generalize well to real-world development scenarios, but it can improve performance when used as an initialization before fine-tuning on real-world data. This finding suggests that LLM-generated simulated and real-world data serve complementary roles rather than interchangeable ones.
\end{itemize}

\section{Related work}
\label{sec:related-work}

\parhead{LLM-based coding assistants}
LLM-based coding assistants have advanced rapidly across software engineering tasks, from code completion and generation~\cite{bruch2009learning,raychev2014code,wang2023review} to program repair~\cite{xia2023automated}, automated test generation~\cite{schafer2023empirical}, and repository-level code comprehension~\cite{ding2023crosscodeeval}. SWE-Agent~\cite{yang2024sweagent} further introduced multi-turn reasoning with tool-use capabilities, enabling models to autonomously navigate codebases, retrieve context, and execute repairs. Meanwhile, commercial coding assistants such as Cursor, GitHub Copilot, and Windsurf have become deeply integrated into the IDE, forming an integral part of users' development workflows. Despite this progress, all existing systems remain fundamentally reactive---providing assistance only upon receiving an explicit user query and unable to intervene when the user's intent has not yet been articulated. Proactive code assistance, the focus of this work, aims to overcome this limitation~\cite{lu2024proactive,chen2025need,zhao2025codinggenie}.

\parhead{Proactive assistance}
Recent research has begun to explore proactive assistance~\cite{deng2023survey,lu2024proactive,kim2025propersim,chen2025need,zhao2025codinggenie}, where the system predicts a user's latent intent from behavioral sequences and context and proactively offers suggestions without an explicit query. ProActiveAgent~\cite{lu2024proactive} first formalized the proactive assistance task and constructed ProActiveBench via LLM-based simulation, covering coding, writing, and other scenarios. ProperSim~\cite{kim2025propersim} further extended the simulation framework to daily-life scenarios. CodingGenie~\cite{zhao2025codinggenie} prototyped a proactive coding assistant within VS Code and evaluated its interaction design via user studies, without systematic benchmarking. A common limitation of the above work is that their training and evaluation data rely entirely on LLM-generated simulated IDE interaction traces~\cite{lu2024proactive,kim2025propersim,zhou2026mind}. Notably, several concurrent efforts~\cite{tang2025proagentbench,chai2026pira} have improved data realism by incorporating real screenshots captured from users' devices. However, screenshots only capture instantaneous interface states and lack fine-grained operational signals---edit deltas, terminal output, cursor movements---as well as repository-level code context. Moreover, these approaches primarily target general Graphical User Interface (GUI) scenarios, making them ill-suited for coding-specific proactive assistance. In contrast, this work presents the first proactive code assistance benchmark derived from real-world IDE interaction traces.

\parhead{Benchmarks and datasets for real-world software development}
A wide range of benchmarks have been established for software engineering research~\cite{li2024deveval,li2024evocodebench}. HumanEval~\cite{chen2021evaluating} and MBPP~\cite{austin2021program} evaluate function-level code generation, SWE-bench~\cite{jimenez2024swebench} extends evaluation to the repository level, and CodeSearchNet~\cite{husain2019codesearchnet} and DevBench~\cite{li2024devbench} provide large-scale data for code comprehension and maintenance. All of these benchmarks, however, take static code or natural-language descriptions as input~\cite{chen2021evaluating,austin2021program,jimenez2024swebench,husain2019codesearchnet,li2024devbench}. Proactive code assistance, by contrast, operates on dynamic IDE interaction traces produced by users during development---a data modality absent from existing software engineering benchmarks. ProCodeBench fills this gap as the first benchmark to incorporate real-world IDE interaction traces into the training and evaluation of proactive code assistance.

\section{Task Definition}
\label{sec:preliminaries}

Unlike reactive coding assistants that rely on explicit instructions, proactive coding assistants aim to predict a developer's implicit intent from the developer's IDE interaction trace and repository context, and to provide assistance before the developer issues an explicit instruction. Formally, we define an IDE operation as a user action recorded during development, such as editing code, switching views, selecting code, or executing a terminal command. Each operation is represented as a tuple $o_i = (p_i, g_i, c_i, t_i)$, where $p_i$ denotes the operation type; $g_i$ denotes the target entity, such as a file, function, or selected code region; $c_i$ denotes the operation content; and $t_i$ denotes the timestamp. A sequence of $n$ consecutive IDE operations forms an IDE interaction trace $\mathcal{O} = \{o_1, o_2, \dots, o_n\}$. Given this trace $\mathcal{O}$ and the repository-level code context $\mathcal{C}$, the task is to predict the developer's intent:
\[
\mathrm{Intent} = f_\theta(\mathcal{O}, \mathcal{C}),
\]
where $f_\theta$ is the proactive coding assistant parameterized by $\theta$. The output is a natural-language intent description. \Cref{fig:example} presents a representative example: after observing that a developer defines a \texttt{retry\_with\_backoff} decorator and then inspects several API-calling modules, the assistant is expected to predict the intent of applying the decorator across the dataflow layer.

\begin{figure}[!t]
\centering
\small
\begin{tcolorbox}[
  colback=white, colframe=tagblue!50, boxrule=0.5pt, arc=3pt,
  left=5pt, right=5pt, top=4pt, bottom=4pt,
  title={\small\textbf{A representative example from ProCodeBench}},
  coltitle=white, colbacktitle=tagblue!70
]
\textbf{Input --- IDE interaction trace $\mathcal{O}$} \textit{(developer working on a Python project)}\\[3pt]
\footnotesize
\begin{tabular}{@{}r@{~}l@{~~}p{6.5cm}@{}}
$o_1$ & \textcolor{tagorange}{\textsc{copy}} & ``\textit{handle API call error and retry; review} \texttt{cli/*.py} \textit{and} \texttt{tradingagents/*.py}''\\
$o_2$ & \textcolor{tagblue}{\textsc{view}}   & open \texttt{dataflows/utils.py}\\
$o_3$ & \textcolor{realgreen}{\textsc{edit}} & in \texttt{utils.py}: add \texttt{import time, functools}; define \texttt{retry\_with\_backoff(retries, backoff)} decorator\\
$o_4$ & \textcolor{tagblue}{\textsc{view}}   & open \texttt{dataflows/alpha\_vantage\_common.py} (contains API functions)\\
$o_5$ & \textcolor{realgreen}{\textsc{edit}} & in \texttt{alpha\_vantage\_common.py}: add \texttt{import time}\\
$o_6$ & \textcolor{tagblue}{\textsc{view}}   & switch back to \texttt{utils.py} to inspect the decorator\\
$o_7$ & \textcolor{tagblue}{\textsc{view}}   & switch back to \texttt{alpha\_vantage\_common.py}\\
\end{tabular}

\tcblower

\textbf{Output --- predicted intent $\mathrm{Intent} = f_\theta(\mathcal{O}, \mathcal{C})$}\\[3pt]
\footnotesize
\textit{Apply the newly defined} \texttt{retry\_with\_backoff} \textit{decorator to API-calling functions across the dataflow modules.}
\end{tcolorbox}
\caption{A representative example from ProCodeBench. The IDE interaction trace---copying a note about API retries, defining a \texttt{retry\_with\_backoff} decorator in \texttt{utils.py}, then opening API-calling modules---naturally suggests the latent intent of \textit{applying the new decorator across the dataflow layer}. Operation types: \textcolor{tagorange}{\textsc{copy}}, \textcolor{tagblue}{\textsc{view}}, \textcolor{realgreen}{\textsc{edit}}.}
\label{fig:example}
\end{figure}

\section{Research methodology}
\label{sec:methodology}

\begin{figure}[!t]
  \centering
  \includegraphics[width=0.85\linewidth]{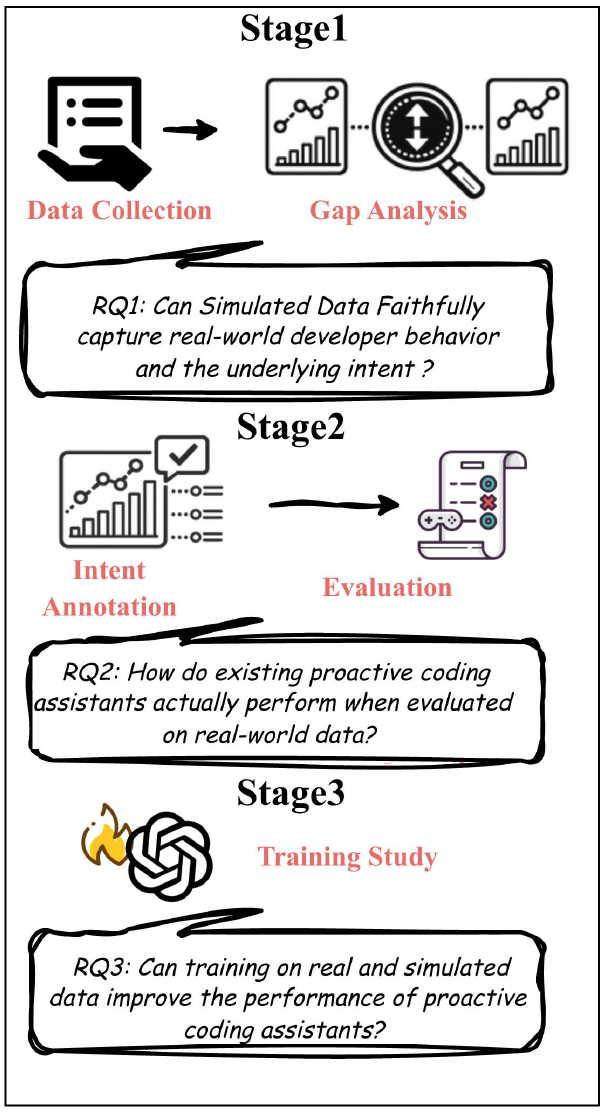}
  \caption{Overview of our research methodology. Stage 1 collects paired real-world and LLM-generated simulated IDE interaction traces for gap analysis. Stage 2 constructs ProCodeBench by annotating real-world traces as intent-prediction instances. Stage 3 compares training regimes based on real-world data, LLM-generated simulated data, and a mixed-data training regime.}
  \label{fig:methodology-overview}
\end{figure}

As shown in \Cref{fig:methodology-overview}, our research methodology comprises three stages, each addressing one research question. \textbf{Stage 1: data collection and gap analysis} collects real-world IDE interaction traces from senior engineer volunteers and pairs each with an LLM-generated simulated counterpart, then quantifies the distributional gap between the two data sources. \textbf{Stage 2: benchmark construction} reconstructs the collected real-world IDE interaction traces into a standardized benchmark for proactive code assistance through an intent annotation pipeline, equipped with a unified evaluation protocol and representative baselines. \textbf{Stage 3: training analysis} investigates the respective roles of real-world and LLM-generated simulated data in model training through a controlled comparison of different training regimes on the benchmark.

\subsection{Stage 1: data collection and gap analysis}
\label{sec:stage1}

To answer RQ1 (\textit{can LLM-generated simulated data faithfully capture real-world developer behavior and the underlying intent?}), we need to obtain large-scale, realistic IDE interaction traces from volunteers, together with paired LLM-generated simulated traces that enable a controlled comparison.

\definecolor{evtHeader}{RGB}{44, 62, 80}      
\definecolor{evtAccent}{RGB}{52, 152, 219}     
\definecolor{evtShadeA}{RGB}{225, 236, 248}    
\definecolor{evtShadeB}{RGB}{240, 244, 248}    
\definecolor{evtRule}{RGB}{180, 190, 200}       
\definecolor{evtRuleLight}{RGB}{205, 215, 225}  

\begin{table}[!tbp]
  \caption{Eight IDE operation types captured by our VS Code extension. Each operation record is stored as a timestamped JSON record with the listed fields.}
  \label{tab:operation-types}
  \centering
  \small
  \setlength{\tabcolsep}{5pt}
  \renewcommand{\arraystretch}{1.35}
  \begin{tabular}{@{} l >{\raggedright\arraybackslash}p{4.8cm} @{}}
    \arrayrulecolor{evtHeader}
    \toprule
    \rowcolor{evtHeader}
    \textbf{\textcolor{white}{Operation type}}
    & \textbf{\textcolor{white}{Captured information}} \\
    \midrule
    \texttt{edit}
    & File path, the inserted and deleted text segments, and the surrounding code context with line range \\
    \arrayrulecolor{evtRule}\hline
    \texttt{copy/paste}
    & Copied or pasted text content from the system clipboard \\
    \hline
    \texttt{view switching}
    & File path, the viewport line range, and the actual code content visible within the viewport \\
    \hline
    \texttt{cursor\_selection}
    & File path, the cursor position or selected text span, and the surrounding code context with a cursor marker \\
    \hline
    \texttt{terminal\_execution}
    & Shell command line, exit code, execution duration, and the captured terminal output \\
    \hline
    \texttt{debug}
    & Debug session identifier, output category (e.g., \texttt{stdout}/\texttt{stderr}), and the captured debug output \\
    \hline
    \texttt{code\_completion}
    & File path, line number, the accepted completion text, and the surrounding code context \\
    \hline
    \texttt{agent\_request}
    & The natural-language request issued by the developer to the coding agent \\
    \arrayrulecolor{evtHeader}
    \bottomrule
  \end{tabular}
\end{table}

\parhead{Real-world data collection}
To collect real-world developer behavior data, we first develop a VS Code extension that records a broad spectrum of IDE operations. As summarized in \Cref{tab:operation-types}, the extension captures eight operation types that cover the most common user-IDE operations, including editing, copy/paste, view switching, cursor selection, terminal execution, and debugging. In particular, it also records AI-assisted development operations, including accepted code completions through \texttt{code completion} and natural-language requests sent to coding agents through \texttt{agent request}. Each operation record is stored as a timestamped JSON record with a structured payload. For example, an \texttt{edit} operation records the edited file path, the inserted and deleted text spans, and the surrounding code context before and after the edit.

Furthermore, we recruit 1{,}246 experienced industry developers as volunteers, covering five major development scenarios: backend development (412, 33.1\%), frontend development (287, 23.0\%), full-stack development (208, 16.7\%), algorithm engineering (183, 14.7\%), and database development (156, 12.5\%), as summarized in \Cref{tab:dataset-stats}. Over three consecutive days, each volunteer works on their own actively maintained industrial project with our VS Code extension enabled. All volunteers receive monetary compensation after completing the collection. In total, we collect approximately \textbf{4.63 million operation events} from real-world IDE interaction traces.

\parhead{LLM-generated simulated data construction}
To construct paired synthetic data, we follow the pipeline of ProActiveAgent~\cite{lu2024proactive}. For each real-world IDE interaction trace, we generate one corresponding LLM-generated simulated trace. To ensure a fair comparison, the simulator is given the same volunteer profile information, including job background, development experience, and primary technology stack. We also constrain each LLM-generated simulated trace to match its paired real-world trace in length and to use the same eight operation types.

\parhead{Gap analysis}
With the paired real-world and LLM-generated simulated IDE interaction traces, we analyze their distributional gap from three perspectives: \textit{behavioral diversity}, \textit{temporal patterns}, and \textit{noise patterns}. \textit{Behavioral diversity} is measured by the frequency distribution of operation types, indicating whether LLM-generated simulated traces cover the long-tail IDE operations observed in real development processes. \textit{Temporal patterns} are characterized through inter-operation time intervals and transition matrices over operation types, revealing whether simulation reproduces the multi-scale pattern of real coding processes and the frequent switches across operation types. \textit{Noise patterns} refer to operations only weakly related to the final intent, such as speculative file browsing, redundant navigation. We further use a representative case study to illustrate how such noise patterns manifest differently in real-world and LLM-generated simulated traces.

\subsection{Stage 2: benchmark construction}
\label{sec:stage2}

\begin{figure}[!t]
  \centering
  \includegraphics[width=\linewidth]{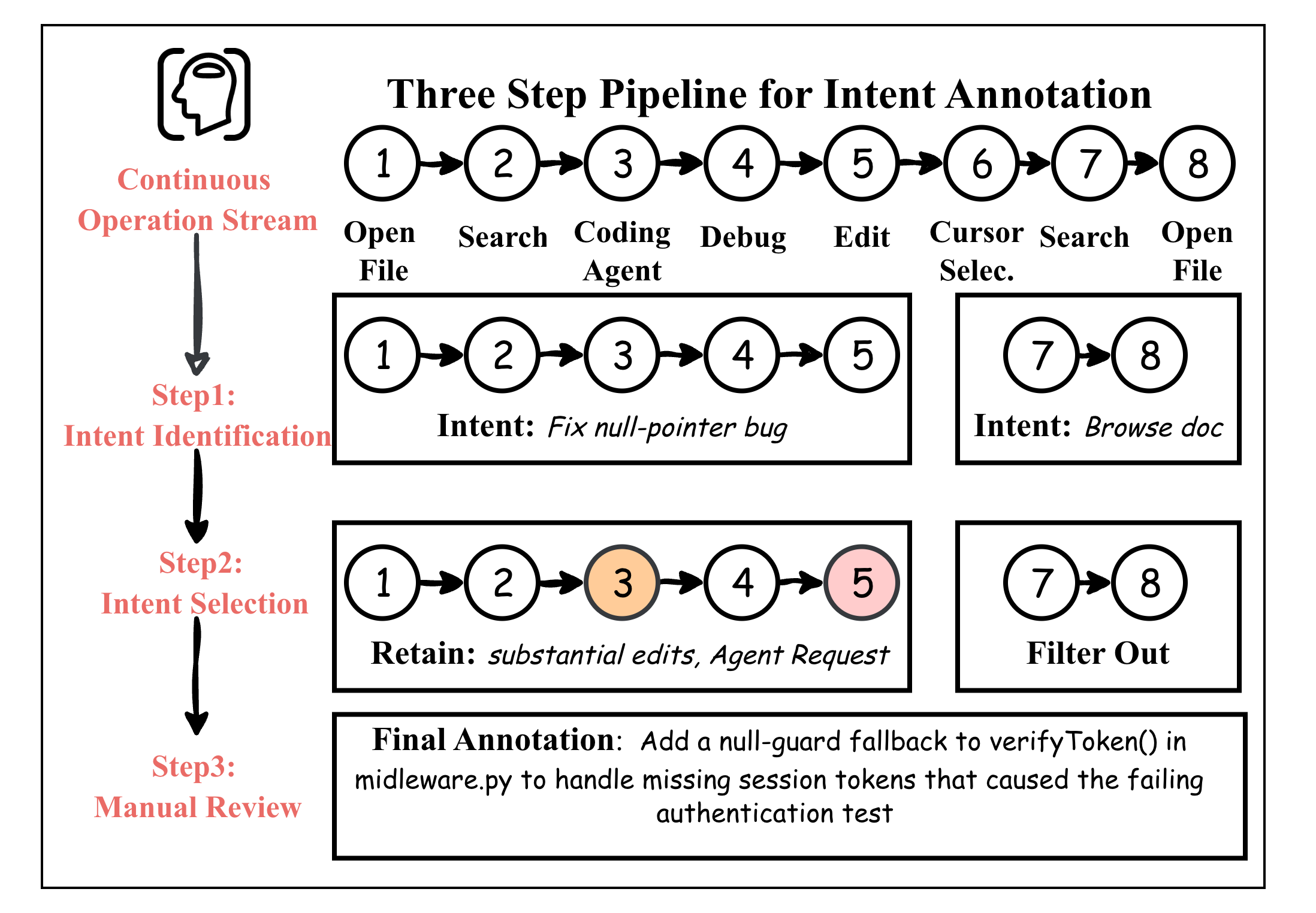}
  \caption{Three-step intent annotation pipeline for converting continuous real-world IDE interaction traces into standardized intent-labeled evaluation samples.}
  \label{fig:annotation-pipeline}
\end{figure}

To answer RQ2 (\textit{how do existing proactive coding assistants perform on real-world data?}), we construct a standardized evaluation benchmark from the collected real-world IDE interaction traces. Unlike LLM-generated simulated data, real-world traces do not come with explicit intent labels. Moreover, developer intent is only implicitly reflected in continuous IDE operations, where a developer may move across different intents, such as refactoring a function or debugging an exception, without clear boundaries between them. Therefore, the main challenge is to convert continuous IDE interaction traces into evaluation instances. To this end, we design a three-step annotation pipeline consisting of intent identification, intent filtering, and manual review (\Cref{fig:annotation-pipeline}). Based on the resulting intent-labeled instances, we build a unified evaluation protocol and a set of representative baselines.

\parhead{Step 1: Intent identification}
To identify developer intents from continuous IDE interaction traces, we adopt a sliding-window strategy. Each window contains $N$ consecutive operations, from which an LLM identifies the latent intents, locates their starting and ending operations, and assigns an initial natural-language intent annotation to each. We set $N=50$ to balance identification quality against LLM annotation cost.

\parhead{Step 2: Intent filtering}
To select evaluation-worthy intent segments, we further filter the candidates identified in Step~1. Since developers do not explicitly specify which intents would be particularly valuable for assistance, we use observable behavioral signals as proxies. We adopt a two-step filtering strategy. The heuristic filtering retains candidate intents with substantial code edits or explicit AI-assistant requests, which often indicate complex development tasks that may benefit from assistance. The semantic filtering uses an LLM to examine whether each retained segment is coherent and consistent with its intent description. 

\parhead{Step 3: Manual review}
To ensure annotation quality, we further conduct a manual review after the automated identification and filtering steps. These automated steps may introduce two types of errors: some retained candidates may have inaccurate intent descriptions, while some valid candidates may have been incorrectly filtered out. Two domain experts independently inspect the candidates, correct erroneous intent descriptions, and recover valid candidates from the filtered set.

\parhead{Dataset split}
After the three-step annotation pipeline, we obtain 5{,}492 valid evaluation samples. To avoid temporal data leakage, we split the samples chronologically into training, validation, and test sets, containing 3{,}576, 1{,}142, and 774 samples, respectively (see \Cref{tab:dataset-stats}). Each evaluation sample takes the preceding IDE interaction trace as input and the corresponding natural-language intent description as output.

\newcommand{\ibar}[3]{\textcolor{#3}{\rule{#1pt}{5pt}}\hspace{1pt}}

\begin{table}[!tbp]
  \caption{ProCodeBench dataset statistics. 1{,}246 volunteers were tracked over 3 consecutive days, yielding 5{,}492 annotated samples.}
  \label{tab:dataset-stats}
  \centering
  \small
  \setlength{\tabcolsep}{4pt}
  \renewcommand{\arraystretch}{1.2}
  \begin{tabular}{@{}l r r l@{}}
    \toprule
    \multicolumn{4}{@{}l}{\cellcolor{tagblue!8}\textbf{\textcolor{tagblue}{Data split}}} \\
    \midrule
    & \textbf{Samples} & \textbf{Ratio} & \\
    Train & 3{,}576 & 65.1\% & \ibar{46}{50}{tagblue} \\
    Validation & 1{,}142 & 20.8\% & \ibar{15}{50}{tagblue!65} \\
    Test & 774 & 14.1\% & \ibar{10}{50}{tagblue!45} \\
    \midrule
    \multicolumn{4}{@{}l}{\cellcolor{tagblue!8}\textbf{\textcolor{tagblue}{Developer domains}}} \\
    \midrule
    & \textbf{Volunteers} & \textbf{Ratio} & \\
    Backend & 412 & 33.1\% & \ibar{41.2}{50}{tagblue} \\
    Frontend & 287 & 23.0\% & \ibar{28.7}{50}{tagblue!75} \\
    Full-stack & 208 & 16.7\% & \ibar{20.8}{50}{tagblue!60} \\
    Algorithm & 183 & 14.7\% & \ibar{18.3}{50}{tagblue!45} \\
    Database & 156 & 12.5\% & \ibar{15.6}{50}{tagblue!35} \\
    \bottomrule
  \end{tabular}
\end{table}

\parhead{Evaluation protocol}
To evaluate the intent-prediction ability of proactive coding assistants, we compare model-generated intents with the developer's real intent. Since developer intents are represented as natural-language descriptions, exact matching is insufficient for evaluating semantic correctness. We therefore adopt an \textbf{LLM-as-a-Judge} evaluation strategy~\cite{zheng2023judging}. For each sample, an independent LLM judge receives the model-generated intent description and the developer's real intent, and determines whether the two are semantically equivalent.

\noindent Unlike prior work that relies entirely on LLM-based judgment, ProCodeBench provides developers' real intents as ground truth, which helps mitigate potential bias from the LLM judge~\cite{wang2024large}. We use \textbf{Pass@$K$} as the primary metric: for each sample, the model independently generates $K$ intent descriptions, and the prediction is considered correct if at least one of them is judged semantically equivalent to the developer's real intent.

\parhead{Baselines}
We compare the following baseline methods on ProCodeBench.

\textit{LLMs.}
To evaluate the intent-prediction ability of current frontier LLMs, we select several widely used models from recent general and software-engineering benchmarks, including DeepSeek-V3.2, GLM-5~\cite{glm5team2026glm5vibecodingagentic}, MiniMax-M2.5, Qwen3.5-397B~\cite{qwen3.5}, GPT-5.4, Claude Sonnet 4.6, and Gemini 3.1 Pro. These models take the developer's IDE interaction trace as input and directly predict the developer's intent.

\textit{Retrieval-Augmented LLMs.}
To evaluate how models use repository-level code context, we include Retrieval-Augmented Generation (RAG)-based methods. Non-graph methods, including RepoCoder~\cite{zhang2023repocoder} and CodeRAG~\cite{zhang2025coderag}, retrieve relevant code fragments through text-based or embedding-based retrieval. Graph-based methods, including GraphCoder~\cite{liu2024graphcoder} and RepoGraph~\cite{ouyang2025repograph}, further incorporate repository structure through static code relations, such as call dependencies, import relations, and symbol-level links.

\textit{LLM-based Agents.}
We also evaluate LLM-based Agents, including SWE-Agent~\cite{yang2024sweagent} and A-RAG~\cite{du2026rag}. Rather than relying on a single retrieval step, these agents obtain repository context through tool-based interaction, such as browsing files, searching code, and inspecting symbols over multiple turns before producing the intent prediction. It is worth noting that previous proactive coding assistants can be viewed as extensions of LLM-based Agents, with an additional capability for predicting user intent proactively. Therefore, we do not treat prior proactive coding assistants as a separate baseline category.

\subsection{Stage 3: training analysis}
\label{sec:stage3}

To answer RQ3 (\textit{can training on simulated or real-world data improve proactive intent prediction?}), we compare the performance of models trained on each data source individually and on the two sources jointly.

\parhead{Training regimes}
We compare three training regimes to study how real-world and LLM-generated simulated data contribute to model training. \textbf{+Real} fine-tunes the backbone on the real-world training set, and \textbf{+Sim.} fine-tunes it on the paired LLM-generated simulated training set. \textbf{+Sim.$\rightarrow$Real} denotes a mixed-data training regime, where the model is first trained on LLM-generated simulated data and then further fine-tuned on real-world data. All regimes are evaluated on the same real-world validation and test sets.

\parhead{Backbone models}
We conduct fine-tuning experiments with open-source models that can be trained under our computational budget of 8$\times$A800-80GB GPUs. To compare training effects across different model families at a similar parameter scale, we select three LLMs: \textbf{Qwen-3-8B}~\cite{qwen3technicalreport}, \textbf{GLM-4-9B}~\cite{glm2024chatglm}, and \textbf{LLaMA-3-8B}~\cite{grattafiori2024llama}.

\section{Results}
\label{sec:results}

In this section, we report experimental results organized around the three research questions.

\subsection{RQ1: Distributional gap between real-world and LLM-generated simulated IDE interaction traces}
\label{sec:gap-analysis}

\looseness=-1
We aim to reveal the distributional gap between real and simulated development data, which constitutes one of the core motivations of ProCodeBench. We analyze this gap from three complementary perspectives: behavioral diversity, temporal patterns, and noise patterns.

\begin{figure}[!t]
  \centering
  \includegraphics[width=\linewidth]{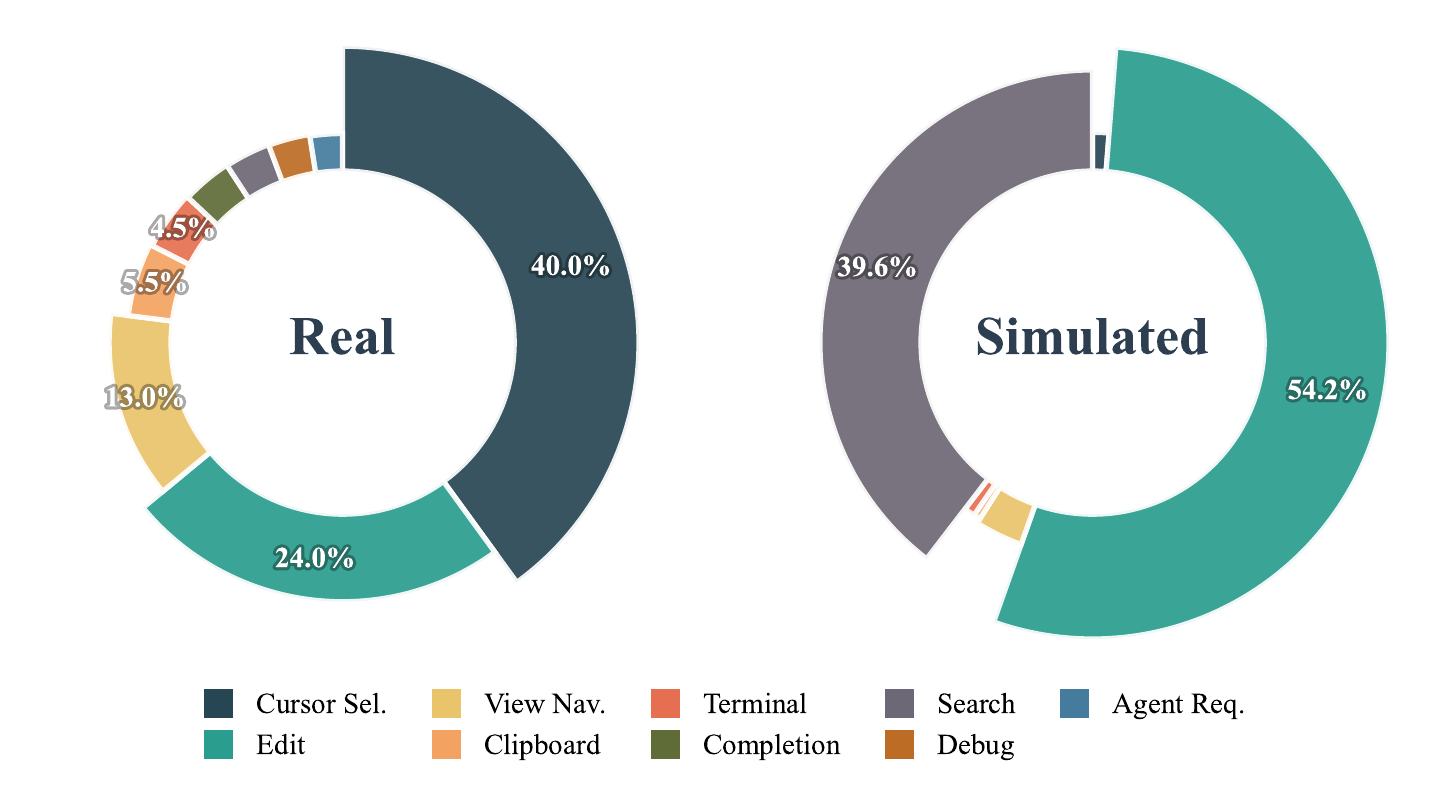}
  \caption{Operation type frequency distribution. Real-world data covers a broader range of operation types with a pronounced long-tail pattern, while LLM-generated simulated data concentrates on a few high-frequency categories.}
  \label{fig:op-type-dist}
\end{figure}

\looseness=-1
\parhead{Behavioral diversity}
As shown in \Cref{fig:op-type-dist}, real-world and LLM-generated simulated traces differ substantially in their operation-type distributions. More specifically, LLM-generated simulated traces are concentrated on a small set of operations, mainly code editing and navigation, whereas real-world traces cover a broader range of IDE operations. Consequently, several operation types that regularly appear in real development processes, including AI-assisted operations, are much less frequent in simulation. Notably, the real-world distribution shows a distinctive pattern: \textbf{cursor selection ($\sim$40\%)} and \textbf{view switching ($\sim$13\%)} account for the largest proportions, rather than code editing. This suggests that developers spend considerable time reading, inspecting, and navigating code before making modifications. Taken together, these observations indicate that LLM-based simulators tend to generate a limited set of operation types, while overlooking the diversity of operations involved in real-world development scenarios.

\begin{figure}[!t]
  \centering
  \includegraphics[width=\linewidth]{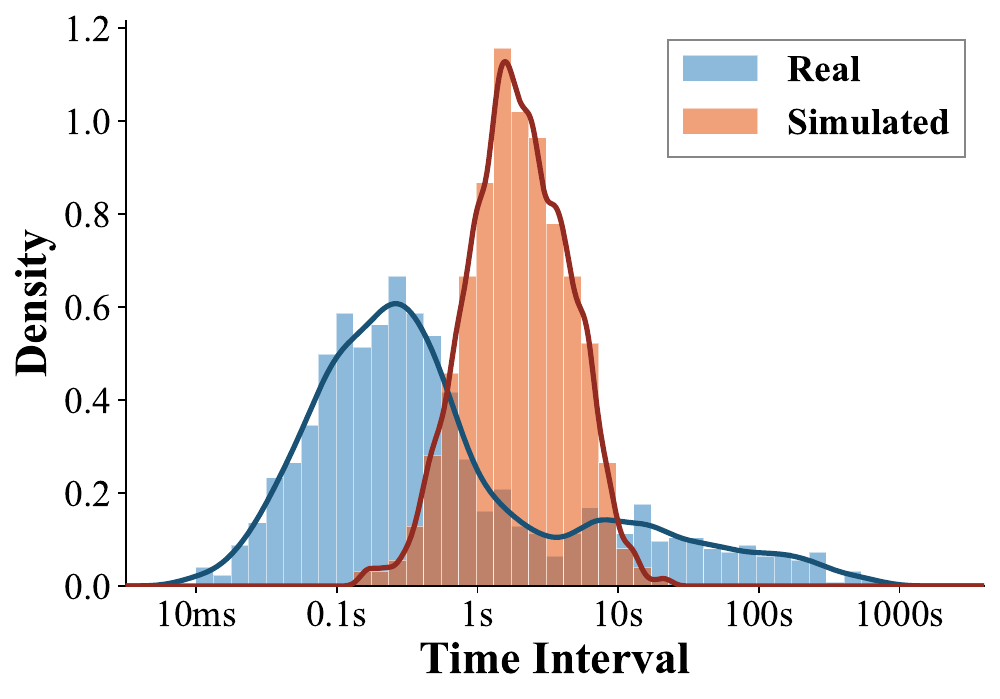}
  \caption{Distribution of inter-operation time intervals. Real-world data exhibits a bimodal pattern with peaks near 0.1s and tens of seconds, while LLM-generated simulated data shows a unimodal distribution centered around 1s.}
  \label{fig:time-interval}
\end{figure}

\begin{figure}[!t]
  \centering
  \includegraphics[width=\linewidth]{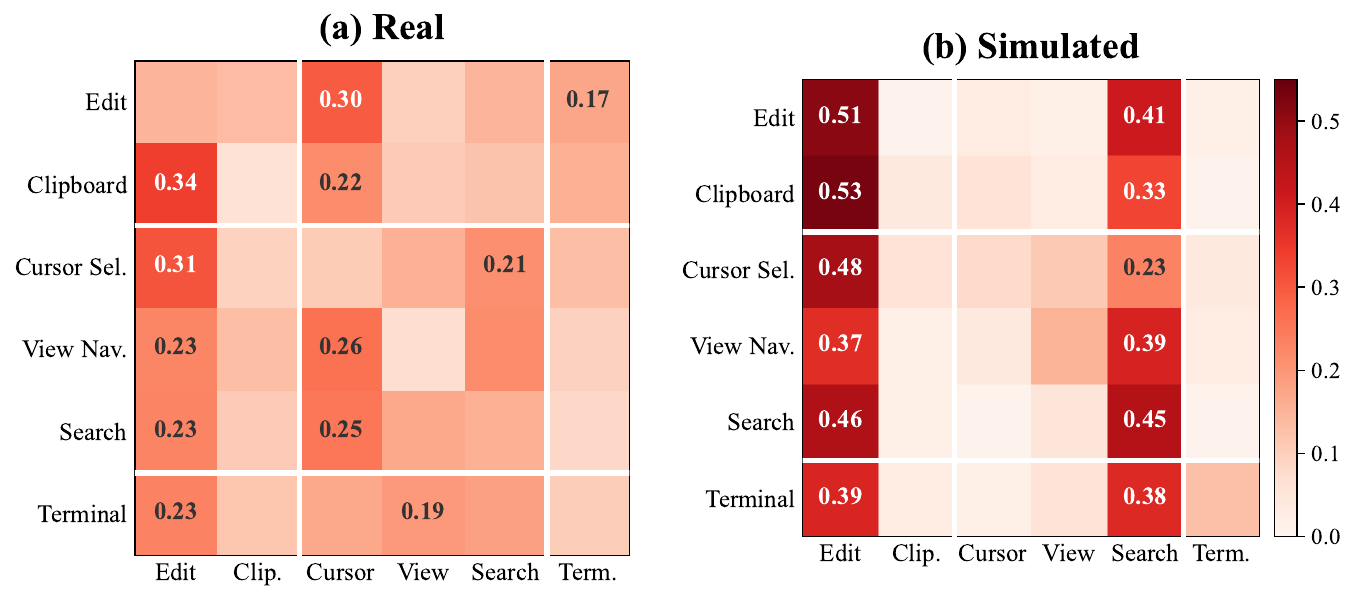}
  \caption{Operation-type transition patterns in real-world and LLM-generated simulated data. Real-world traces involve diverse transitions across operation types, whereas LLM-generated simulated traces concentrate on a small set of frequent transitions.}
  \label{fig:transition}
\end{figure}

\looseness=-1
\parhead{Temporal patterns}
To reveal how real-world and LLM-generated simulated traces differ in temporal patterns, we analyze both inter-operation intervals and operation-type transitions. For inter-operation intervals, real-world traces exhibit a pronounced bimodal distribution, with one peak near 0.1\,s and another around tens of seconds (\Cref{fig:time-interval}). This pattern reflects the multi-scale temporal structure of real development, where short bursts of operations are interleaved with longer pauses. By contrast, LLM-generated simulated traces show a unimodal distribution centered around 1\,s, indicating a more regular temporal structure. For operation-type transitions, real-world traces spread across many pairs of operation types (\Cref{fig:transition}), suggesting frequent switches among different development activities. However, LLM-generated simulated traces concentrate on a small number of adjacent transition pairs, such as Edit$\to$Edit and Edit$\to$Navigation. Taken together, these results indicate that LLM-based simulators do not adequately reproduce the temporal structure and operation-switching patterns of real coding processes.

\looseness=-1
\parhead{Noise patterns}
Real-world traces contain many operations that are only weakly related to the final intent, since developers often adjust their actions while exploring and refining their solution. As illustrated in \Cref{fig:noise-case}, these operations may include irrelevant browsing, repeated navigation, or reverted edits. Although they introduce noise patterns, they also reflect the uncertainty and nonlinearity of real development processes. By contrast, LLM-generated simulated traces rarely contain similar exploratory or corrective behavior. Instead, they tend to generate simplified action sequences. Taken together, this contrast indicates that real-world traces contain human exploratory noise, whereas LLM-generated simulated traces exhibit an oversimplification bias.

\begin{boxK}
\small \textbf{Key finding \ding{182}:} \textit{LLM-generated simulated traces fail to faithfully capture real-world developer behavior. Compared with real-world IDE interaction traces, LLM-generated simulated traces show reduced behavioral diversity, more regular temporal patterns, and omit much of the exploratory behavior present in real coding processes.}
\end{boxK}

\begin{figure}[!t]
\centering
\small
\begin{tcolorbox}[
  colback=white, colframe=black!60, boxrule=0.4pt, arc=2pt,
  left=4pt, right=4pt, top=3pt, bottom=3pt,
  title={\small\textbf{Case study: noise patterns in real-world vs.\ LLM-generated simulated IDE interaction traces}},
  coltitle=white, colbacktitle=black!60
]
\textbf{Real-world IDE interaction trace} (intent: \textit{fix null-pointer crash in login handler})\\[2pt]
{\footnotesize
\texttt{[1]} \textcolor{gray}{open} \texttt{auth.js} \hfill \textcolor{realgreen}{\ding{51}}\\
\texttt{[2]} \textcolor{gray}{navigate to} \texttt{handleLogin()} \hfill \textcolor{realgreen}{\ding{51}}\\
\texttt{[3]} \textcolor{gray}{set breakpoint} \texttt{auth.js:42} \hfill \textcolor{realgreen}{\ding{51}}\\
\texttt{[4]} \textcolor{gray}{run} \texttt{npm run dev} \hfill \textcolor{realgreen}{\ding{51}}\\
\texttt{[5]} \textcolor{gray}{open} \texttt{config.js} \hfill \textcolor{missred}{\ding{55}\,redundant}\\
\texttt{[6]} \textcolor{gray}{open} \texttt{README.md} \hfill \textcolor{missred}{\ding{55}\,redundant}\\
\texttt{[7]} \textcolor{gray}{back to} \texttt{auth.js} \hfill \textcolor{missred}{\ding{55}\,redundant}\\
\texttt{[8]} \textcolor{gray}{navigate to} \texttt{userService.js} $\to$ \texttt{getUser()} \hfill \textcolor{realgreen}{\ding{51}}\\
\texttt{[9]} \textcolor{gray}{back to} \texttt{auth.js} \hfill \textcolor{missred}{\ding{55}\,redundant}\\
\texttt{[10]} \textcolor{gray}{edit} \texttt{handleLogin()}: add null guard \hfill \textcolor{realgreen}{\ding{51}}\\
\texttt{[11]} \textcolor{gray}{undo} last edit \hfill \textcolor{missred}{\ding{55}\,redundant}\\
\texttt{[12]} \textcolor{gray}{edit} \texttt{handleLogin()}: add guard with fallback \hfill \textcolor{realgreen}{\ding{51}}\\
\texttt{[13]} \textcolor{gray}{run} \texttt{npm test} \hfill \textcolor{realgreen}{\ding{51}}\\
\texttt{[14]} \textcolor{gray}{select} test output: \texttt{1 failed} \hfill \textcolor{realgreen}{\ding{51}}\\
\texttt{[15]} \textcolor{gray}{edit} \texttt{handleLogin()}: fix edge case \hfill \textcolor{realgreen}{\ding{51}}\\
\texttt{[16]} \textcolor{gray}{run} \texttt{npm test} \hfill \textcolor{realgreen}{\ding{51}}
}
\tcblower
\textbf{LLM-generated simulated IDE interaction trace} (intent: \textit{add retry logic to API request})\\[2pt]
{\footnotesize
\texttt{[1]} \textcolor{gray}{open} \texttt{api.js} \hfill \textcolor{realgreen}{\ding{51}}\\
\texttt{[2]} \textcolor{gray}{navigate to} \texttt{fetchData()} \hfill \textcolor{realgreen}{\ding{51}}\\
\texttt{[3]} \textcolor{gray}{edit} \texttt{fetchData()}: add retry wrapper \hfill \textcolor{realgreen}{\ding{51}}\\
\texttt{[4]} \textcolor{gray}{edit} \texttt{fetchData()}: add max retry config \hfill \textcolor{realgreen}{\ding{51}}\\
\texttt{[5]} \textcolor{gray}{open} \texttt{api.test.js} \hfill \textcolor{realgreen}{\ding{51}}\\
\texttt{[6]} \textcolor{gray}{edit} \texttt{api.test.js}: add retry test case \hfill \textcolor{realgreen}{\ding{51}}\\
\texttt{[7]} \textcolor{gray}{run} \texttt{npm test} \hfill \textcolor{realgreen}{\ding{51}}
}
\end{tcolorbox}
\caption{Representative noise patterns. The real-world IDE interaction trace contains exploratory browsing (\texttt{[3--4]}), redundant navigation (\texttt{[5]}), and a reverted edit (\texttt{[7]}), whereas the LLM-generated simulated IDE interaction trace proceeds linearly without trial-and-error.}
\label{fig:noise-case}
\end{figure}

\subsection{RQ2: Performance of existing proactive coding assistants on real-world data}
\label{sec:baseline-comparison}

\looseness=-1
\parhead{LLM results}
As shown in \Cref{tab:main-results}, all LLMs achieve limited performance on the real-world test set, with Pass@1 below 14\% across all models. Even the strongest model, Claude Sonnet 4.6, reaches only 13.57\%. This result indicates that current frontier LLMs have limited ability to predict developer intent from real-world IDE interaction traces. Compared with the substantially stronger results reported on simulation-based benchmarks~\cite{lu2024proactive,kim2025propersim}, the results further suggest that simulation-based evaluation may overestimate real-world intent-prediction performance.

We also observe that model rankings on ProCodeBench do not align with those commonly observed on general software-engineering benchmarks~\cite{jimenez2024swebench}. For instance, GPT-5.4 achieves 6.59\% Pass@1, lagging behind Claude Sonnet 4.6 (13.57\%) and Gemini 3.1 Pro (11.46\%). This suggests that proactive intent prediction requires capabilities beyond general reasoning, and improving this ability remains an open challenge.

\looseness=-1
\parhead{Retrieval-Augmented LLM and LLM-based Agent results}
Across all backbone models, Retrieval-Augmented LLMs and LLM-based Agents outperform the corresponding LLMs, indicating that IDE interaction traces alone are often insufficient for proactive intent prediction. For example, with GLM-5 as the backbone, Pass@1 increases from 7.77\% to 9.49\%--16.73\% after repository context is introduced. Similar gains are observed for GPT-5.4, DeepSeek-V3.2, and Qwen3.5.

The improvement is especially pronounced for LLM-based Agents. Compared with Retrieval-Augmented LLMs, LLM-based Agents achieve the highest Pass@1 across all four backbones. For instance, A-RAG reaches 16.73\% with GLM-5, 35.57\% with GPT-5.4, 15.81\% with DeepSeek-V3.2, and 15.02\% with Qwen3.5. This indicates that autonomous multi-turn tool use is important for proactive intent prediction.

However, this performance gain comes with a substantial efficiency cost. LLM-based Agents require an average of 23 tool interactions per prediction, which increases computation and response latency. Developing methods that achieve both stronger performance and improved efficiency remains an important direction for future proactive code assistance.

\begin{boxK}
\small \textbf{Key finding \ding{183}:} \textit{Existing models perform poorly on real-world benchmark, suggesting that simulation-based benchmarks may overestimate their intent-prediction ability. Repository-level code context consistently improves performance, but obtaining useful context efficiently remains a key challenge.}
\end{boxK}

\begin{table}[!tbp]
  \centering
  \small
  \caption{LLM Pass@$K$ accuracy (\%) on ProCodeBench (759 samples). \best{Red}: best, \second{Blue}: runner-up.}
  \label{tab:main-results}
  \setlength{\tabcolsep}{6pt}
  \renewcommand{\arraystretch}{1.15}
  \begin{tabular}{lccc}
    \toprule
    \textbf{Model} & \textbf{Pass@1} & \textbf{Pass@3} & \textbf{Pass@5} \\
    \midrule
    Claude-Sonnet-4-6 & \best{13.57} & \best{21.61} & \best{24.37} \\
    \rowcolor{gray!5}
    Gemini-3.1-Pro & \second{11.46} & \second{17.79} & \second{21.87} \\
    Qwen3.5-397B & 8.43 & 15.68 & 16.21 \\
    \rowcolor{gray!5}
    GLM-5 & 7.77 & 12.65 & 15.42 \\
    DeepSeek-V3.2 & 7.77 & 13.44 & 19.24 \\
    \rowcolor{gray!5}
    GPT-5.4 & 6.59 & 16.34 & 19.10 \\
    MiniMax-M2.5 & 2.77 & 6.46 & 7.91 \\
    \bottomrule
  \end{tabular}
\end{table}

\begin{table*}[!t]
  \centering
  \small
  \caption{Pass@1 accuracy (\%) of Retrieval-Augmented LLMs and LLM-based Agents on ProCodeBench (759 samples). \best{Red}: best, \second{Blue}: runner-up per row.}
  \label{tab:agent-results}
  \setlength{\tabcolsep}{5pt}
  \renewcommand{\arraystretch}{1.15}
  \begin{tabular}{l|cccc|cc}
    \toprule
    \multirow{2}{*}{\textbf{Model}} &
      \multicolumn{4}{c|}{\textbf{Retrieval-Augmented LLMs}} &
      \multicolumn{2}{c}{\textbf{LLM-based Agents}} \\
    & RepoCoder & CodeRAG & GraphCoder & RepoGraph & SWE-Agent & A-RAG \\
    \midrule
    GLM-5 & 9.49 & 10.01 & 11.07 & 10.54 & \second{14.23} & \best{16.73} \\
    \rowcolor{gray!5}
    GPT-5.4 & 10.67 & 11.46 & 14.36 & 13.57 & \second{32.98} & \best{35.57} \\
    DeepSeek-V3.2 & 9.88 & 10.41 & 11.86 & 11.33 & \second{12.52} & \best{15.81} \\
    \rowcolor{gray!5}
    Qwen3.5 & 10.14 & 10.67 & 11.59 & 11.07 & \second{12.52} & \best{15.02} \\
    \bottomrule
  \end{tabular}
\end{table*}

\begin{table}[!tbp]
  \centering
  \small
  \caption{Pass@1 accuracy (\%) across four training regimes. \best{Red}: best per column.}
  \label{tab:real-vs-sim}
  \setlength{\tabcolsep}{5pt}
  \renewcommand{\arraystretch}{1.15}
  \begin{tabular}{lccc}
    \toprule
    \textbf{Regime} & \textbf{Qwen-3-8B} & \textbf{GLM-4-9B} & \textbf{LLaMA3-8B} \\
    \midrule
    Backbone & 2.53 & 2.24 & 1.84 \\
    \rowcolor{gray!5}
    +Sim. & 1.97 & 1.65 & 1.32 \\
    +Real & 5.84 & 4.93 & 3.76 \\
    \rowcolor{gray!5}
    +Sim.$\rightarrow$Real & \best{7.63} & \best{6.52} & \best{5.21} \\
    \bottomrule
  \end{tabular}
\end{table}

\subsection{RQ3: Training study}
\label{sec:training-analysis}

\begin{figure}[!t]
  \centering
  \includegraphics[width=\linewidth]{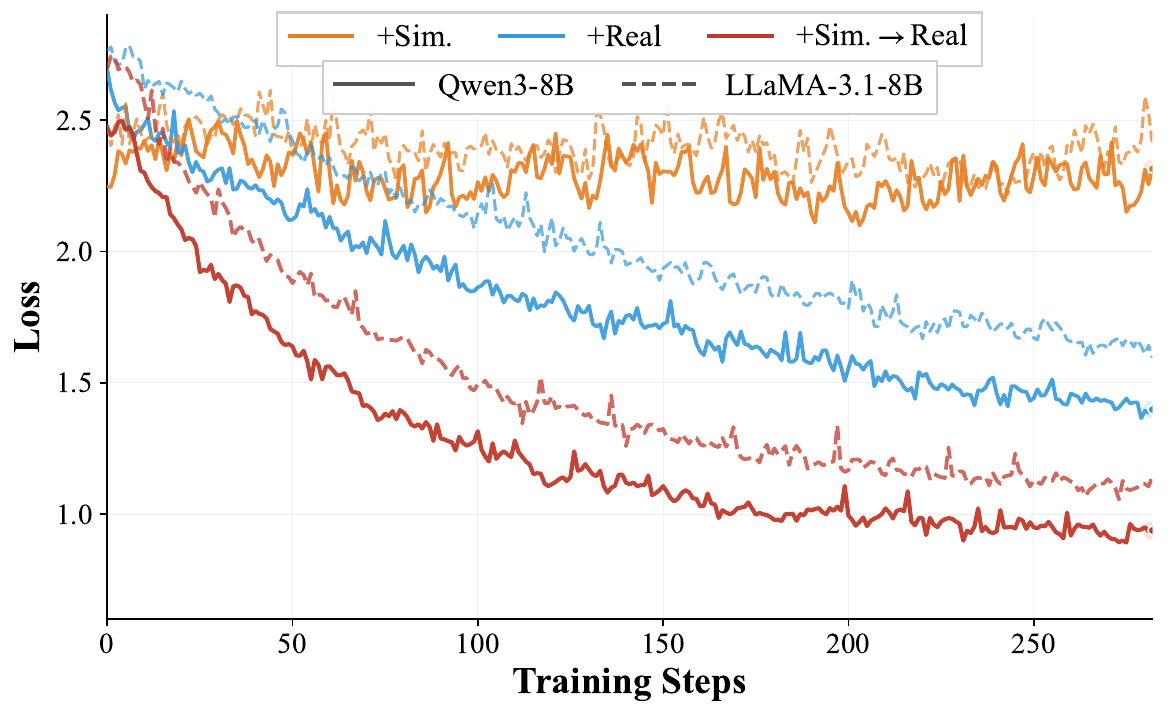}
  \caption{Training loss curves during fine-tuning on real-world data under different training regimes. +Sim.$\rightarrow$Real converges significantly faster and reaches a lower final loss than +Real.}
  \label{fig:training-loss}
\end{figure}

\looseness=-1
To answer RQ3, we compare three training regimes on three open-source backbones: Qwen-3-8B, GLM-4-9B, and LLaMA3-8B. The three regimes include fine-tuning on LLM-generated simulated data only (+Sim.), fine-tuning on real-world data only (+Real), and a mixed-data training regime (+Sim.$\rightarrow$Real). All regimes are evaluated on the same real-world test set with Pass@1. We also track the training loss to examine the optimization behavior of different regimes. Results are reported in \Cref{tab:real-vs-sim} and \Cref{fig:training-loss}.

\looseness=-1
\parhead{Single-source training}
As shown in \Cref{tab:real-vs-sim}, real-world data provides a clear benefit for model training, whereas LLM-generated simulated data alone does not transfer effectively to the real-world test set. Fine-tuning on real-world data (+Real) consistently improves all three backbones, increasing Pass@1 from 2.53\% to 5.84\% for Qwen-3-8B, from 2.24\% to 4.93\% for GLM-4-9B, and from 1.84\% to 3.76\% for LLaMA3-8B. In contrast, fine-tuning on LLM-generated simulated data alone (+Sim.) reduces performance below the original backbone, with Pass@1 dropping to 1.97\%, 1.65\%, and 1.32\%, respectively. These results suggest that the gap between LLM-generated simulated and real-world traces directly affects training: models trained only on LLM-generated simulated data fail to generalize to real-world proactive intent prediction.

\looseness=-1
\parhead{Complementarity between simulated and real-world data}
Although LLM-generated simulated data performs poorly as a standalone training source, it becomes useful when followed by real-world fine-tuning. The mixed-data training regime (+Sim.$\rightarrow$Real) achieves the best Pass@1 across all three backbones, reaching 7.63\% on Qwen-3-8B, 6.52\% on GLM-4-9B, and 5.21\% on LLaMA3-8B. Compared with +Real, this corresponds to gains of 1.79, 1.59, and 1.45 percentage points, respectively. The training loss curves in \Cref{fig:training-loss} further show that models initialized with LLM-generated simulated data converge faster and reach a lower final loss during real-world fine-tuning. These results suggest that LLM-generated simulated data can provide a useful warm start, while real-world data is still necessary to adapt the model to real-world development scenarios. Therefore, LLM-generated simulated and real-world data should not be viewed as interchangeable sources, but as complementary signals for training proactive coding assistants.

\begin{boxK}
\small \textbf{Key finding \ding{184}:} \textit{Although LLM-generated simulated data cannot substitute for real-world data, the two complement each other---their combination yields performance gains beyond either source alone.}
\end{boxK}

\section{Discussion}
\label{sec:discussion}

\subsection{Ablation study}
\label{sec:ablation}

\looseness=-1
To understand how different operation types contribute to proactive intent prediction, we conduct an operation-type ablation study. Starting from the full IDE interaction trace, we remove one operation type at a time and measure the resulting change in Pass@1. To examine whether different models use these signals in different ways, we evaluate three models with different performance levels: Claude Sonnet 4.6, Gemini 3.1 Pro, and MiniMax-M2.5. Results are reported in \Cref{tab:ablation-ops}.

\begin{table}[!tbp]
  \centering
  \small
  \caption{Impact of operation types on Pass@1 (\%). Each row removes one operation type (\ding{55}). $\downarrow$: degradation, $\uparrow$: improvement relative to the full IDE interaction trace.}
  \label{tab:ablation-ops}
  \setlength{\tabcolsep}{3.5pt}
  \renewcommand{\arraystretch}{1.15}
  \begin{tabular}{lccc}
    \toprule
    \textbf{Setting} & \textbf{Claude 4.6} & \textbf{Gemini 3.1} & \textbf{MiniMax} \\
    \midrule
    Full sequence & 13.57 & 11.46 & 2.77 \\
    \midrule
    \multicolumn{4}{l}{\textit{Navigation \& Selection}} \\
    \rowcolor{gray!5}
    \ding{55}\,cursor\_sel. & 12.78\,{\scriptsize$\downarrow$0.79} & 11.20\,{\scriptsize$\downarrow$0.26} & 3.16\,{\scriptsize$\uparrow$0.39} \\
    \ding{55}\,copy/paste & 13.04\,{\scriptsize$\downarrow$0.53} & 11.20\,{\scriptsize$\downarrow$0.26} & 2.90\,{\scriptsize$\uparrow$0.13} \\
    \rowcolor{gray!5}
    \ding{55}\,view switching & 12.52\,{\scriptsize$\downarrow$1.05} & 10.67\,{\scriptsize$\downarrow$0.79} & 3.03\,{\scriptsize$\uparrow$0.26} \\
    \midrule
    \multicolumn{4}{l}{\textit{Execution \& Editing}} \\
    \ding{55}\,terminal & 11.86\,{\scriptsize$\downarrow$1.71} & 9.88\,{\scriptsize$\downarrow$1.58} & 2.37\,{\scriptsize$\downarrow$0.40} \\
    \rowcolor{gray!5}
    \ding{55}\,edit & 4.22\,{\scriptsize$\downarrow$9.35} & 3.16\,{\scriptsize$\downarrow$8.30} & 0.66\,{\scriptsize$\downarrow$2.11} \\
    \midrule
    \multicolumn{4}{l}{\textit{AI Interaction}} \\
    \ding{55}\,agent\_req. & 13.31\,{\scriptsize$\downarrow$0.26} & 11.33\,{\scriptsize$\downarrow$0.13} & 2.64\,{\scriptsize$\downarrow$0.13} \\
    \bottomrule
  \end{tabular}
\end{table}

\looseness=-1
\parhead{Comparison across operation types}
As shown in \Cref{tab:ablation-ops}, removing \texttt{edit} causes the largest performance drop across all three models, decreasing Pass@1 by 9.35, 8.30, and 2.11 percentage points, respectively. Removing \texttt{terminal\_execution} also leads to degradation. These results indicate that code edits and execution feedback provide the most direct information for predicting developer intent. In contrast, removing \texttt{agent\_request} has only a minor effect, likely because this operation type appears less frequently. Overall, current models rely heavily on edit and execution signals, while many other behavioral signals in real-world IDE interaction traces remain underutilized.

\looseness=-1
\parhead{Sensitivity across models}
Models with different capabilities show different sensitivity to operation-type removal. Claude Sonnet 4.6 degrades under every ablation setting, suggesting that it can use a broader range of IDE signals. In particular, navigation and selection operations provide complementary context for stronger models. By contrast, MiniMax-M2.5 slightly improves when \texttt{cursor\_selection}, \texttt{copy/paste}, or \texttt{view switching} is removed, suggesting that less capable models may treat these signals as noise rather than useful information.

\begin{boxK}
\small \textbf{Key finding \ding{185}:} \textit{Current models rely heavily on explicit execution information, while stronger models can also use contextual signals from other operations. This suggests that effectively leveraging developers' IDE interaction traces remains a key challenge for proactive intent prediction.}
\end{boxK}

\subsection{Threats to validity}
\label{sec:threats-validity}

\parhead{Limitations of intent annotation}
Developer intent in real-world IDE interaction traces is implicit and cannot be directly observed. Therefore, our annotation pipeline relies on an LLM and observable behavioral signals, such as substantial code edits and explicit AI-assistant requests, to infer candidate intents. This process may miss some intents that would be valuable for proactive assistance. To mitigate this threat, domain experts independently review the candidate intents in the final stage of the pipeline, correct inaccurate intent, and recover valid candidates that were incorrectly filtered out. Nevertheless, accurately recovering developer intent from IDE interaction traces remains an open challenge.

\parhead{Data collection scope}
Limited by the cost and privacy risks of collecting real-world IDE interaction traces, it is impractical to exhaustively cover all development time scales and developer populations. We mitigate this threat by recruiting 1{,}246 experienced industry developers from five major development scenarios, including frontend, backend, full-stack, database, and algorithm engineering, and by collecting traces from their own active projects. This design provides broad coverage of real-world daily development behavior.

\parhead{Restricted release for privacy}
The IDE interaction traces in ProCodeBench may contain sensitive commercial information, such as proprietary enterprise repositories. Although we obtained consent from volunteers through a data collection agreement, the agreement does not permit unrestricted public release of the collected data. Therefore, we cannot fully open-source the raw dataset. To support reproducibility, we instead provide a controlled-access evaluation platform for academic research. Researchers may apply for access with institutional information; once approved, they can submit their model or agent system and obtain evaluation results on ProCodeBench.

\section{Conclusion}
\label{sec:conclusion}

\looseness=-1
This paper investigates whether LLM-generated simulated data can faithfully support proactive code assistance in real-world development scenarios. To analyze this question, we first collect large-scale real-world IDE interaction traces from 1,246 volunteers through a VS Code extension, and pair each real-world trace with an LLM-generated simulated counterpart. This paired design allows us to directly examine the simulation-to-reality gap in developer behavior. Our analysis shows that LLM-generated simulated traces differ substantially from real-world traces in behavioral diversity, temporal structure, and exploratory operations, indicating that simulation alone cannot fully capture how developers work in practice.

\looseness=-1
Building on the collected real-world data, we construct ProCodeBench, a standardized benchmark for proactive intent prediction in real-world development scenarios. Experiments on LLMs, Retrieval-Augmented LLMs, and LLM-based Agents show that existing models still struggle on this benchmark, suggesting that simulation-based evaluation may overestimate intent-prediction ability. We further study how real-world and LLM-generated simulated data affect training. The results show that LLM-generated simulated data alone does not transfer well to real-world development scenarios, but can provide useful initialization before training on real-world data. These findings suggest that future proactive coding assistants should be evaluated on real-world developer behavior and should learn to use the rich but noisy signals contained in IDE interaction traces.

\section*{Acknowledgment}
\noindent


\bibliographystyle{ieeetr}
\bibliography{doc}

\end{document}